\documentclass[12pt]{iopart}

\usepackage[dvips]{graphicx}
\usepackage[dvips]{color}
\usepackage{hyperref}
\usepackage{float}

\usepackage{iopams}

\usepackage{setstack}

\usepackage{float}

\providecommand{\openone}{\leavevmode\hbox{\small1\kern-3.8pt\normalsize1}}

\newcommand{\Trace}{\mathrm{Tr}}
\newcommand{\Real}{\mathrm{Re}}

\newcommand{\bra}{\langle}
\newcommand{\ket}{\rangle}

\newcommand{\mmax}[2]{\ifdim#1>#2#1\else#2\fi}
\newcommand{\mmin}[2]{\ifdim#1<#2#1\else#2\fi}

\newcommand{\Ket}[1]{|#1\rangle}

\begin{document}
\title[On-site Spin-orbit for LCAO]{On-site approximation for spin-orbit coupling in LCAO density functional methods}
\author{L Fern\'andez-Seivane,$^{1,*}$ M A Oliveira,$^{2}$ S Sanvito$^2$ and J Ferrer$^1$}
\address{$^1$ Departamento de F\'{\i}sica, Universidad de Oviedo,
33007 Oviedo, Spain}
\address{$^2$ Department of Physics, Trinity
College, Dublin 2, Ireland}
\ead{$^*$ quevedin@condmat.uniovi.es}
\date{\today}

\message{\the\linewidth}

\begin{abstract}
We propose a computational method that simplifies drastically the
inclusion of spin-orbit interaction in density functional theory when
implemented over localised atomic orbital basis sets.
Our method is based on a well-known procedure for obtaining pseudopotentials
from atomic relativistic {\it ab initio} calculations and on an on-site approximation
for the spin-orbit matrix elements. We have implemented the technique in the
SIESTA\cite{siesta} code, and show that it provides accurate
results for the overall band structure and splittings of group {IV} and {III-IV}
semiconductors as well as for 5$d$ metals.
\end{abstract}
\maketitle
\JPCM 18 7999-8013, 2006
\section{Introduction}
\label{intro}
Computational methods in condensed matter physics are a
powerful tool for predicting and explaining the most diverse properties of
materials, nanostructures and small biological systems.
Among an enormous plethora of methodologies, density functional
theory (DFT) \cite{DFT,LDA} has become a standard for simulations at the atomic and
nanometric scale. Several numerical implementations of DFT are available and
the details of the algorithm usually depends on the specific applications
the method is designed for. These implementations can be categorized according to two
different schemes.

The first scheme divides DFT codes depending on the basis functions used, namely plane-waves
or linear combination of atomic orbitals (LCAO). Plane waves are typically easier to implement and
the convergence is determined by a single variational parameter, the energy cutoff. In contrast LCAO
implementations are based on a tight binding description of the chemical bond. They are more
cumbersome to implement and the variational principle is controlled by a set of parameters defining
the basis functions. However these second methods are ideal for linear scaling since sparse Hamiltonian
can be constructed by using strictly confined orbitals \cite{orderN}.

The second classification takes into account whether the codes simulate both core and
valence electrons or only the valence ones. In the first case the method is regarded as
{\it all electron} since all the electronic degrees of freedom are treated on the same footing. This
for instance allows one to perform fully relativistic calculations without any conceptual
complication. All electron methods are remarkably accurate but have the drawback that the
calculations are usually rather intensive and only relatively small systems can be tackled.
In the second case the contribution of the core electrons is casted into pseudopotentials \cite{martin},
which also can be constructed from DFT. This reduces drastically the number of electrons considered
in the self-consistent simulation and much larger systems can be investigated.

Spin-orbit interaction is a relativistic effect whose magnitude increases with the atomic number.
Consequently it provides negligible contributions to the electronic structure of individual atoms and bulk
materials made of light elements. However it has a significant impact over the physics of heavier elements,
for instance 3$d$ ferromagnetic materials. Spin-orbit can produce magnetic anisotropies of the order of
10 to 100 $\mu$eV for bcc and fcc Fe, Ni and Co \cite{suecos}, therefore is the underlying mechanism
for establishing the magnetic easy and hard axes and for controlling the shape of domain walls \cite{bruno}.
It is also the primary interaction responsible for most of the zero field splitting and other properties of
magnetic molecules \cite{roberta}.

In semiconductors spin-orbit interaction spin-splits the edges of the valence and the conduction
band \cite{cardona} and allows electrical manipulation of the spin-direction \cite{dasdatta}.
This last effect is of paramount importance for the growing field of spintronics \cite{spintronics},
which has certainly added more impetus to the inclusion of the spin-orbit interaction in the
description of the electronic structure. Spin-orbit coupling determines the spin-relaxation time of
electrons in ordinary semiconductors and in semiconductor heterostructures, \cite{stime} and also
plays an important role in the physics of diluted magnetic semiconductors. \cite{DMS} Finally is worth
mentioning that electron spin manipulation using spin-orbit interaction was recently demonstrated
in the so-called spin-Hall effect \cite{spinhall}, a solid state version of the Stern-Gerlach measurement.

It is therefore clear that spin-orbit interaction is becoming increasingly important for a number of applications,
which also require the description of rather large systems. This calls for an efficient implementation
of spin-orbit within pseudopotential LCAO based algorithms. Interestingly most of the mainstream LCAO codes
such as SIESTA \cite{siesta}, Onetep \cite{onetep}, Fireball \cite{fireball} and Conquest \cite{conquest} do not
contain implementations of spin-orbit interaction in their present form. In contrast quantum chemistry packages
such as Gaussian\cite{gaussian} or Turbomole \cite{turbomole} are not equipped for solid-state simulations.

Therefore we have developed a general method for including spin-orbit interaction in conventional pseudopotential
based LCAO DFT methods. This is not computationally demanding and hence it is fully adequate for large scale
simulations. The method, although not suitable for highly accurate total energy calculations for which all electron
plane-wave schemes cannot be matched, provides an excellent description of the effects of spin-orbit coupling on
the electronic structure. Here we describe our implementation within the SIESTA program, \cite{siesta}
although the scheme is very general and it could be readily implemented in any LCAO pseudopotential codes
with non-collinear electron spin functionalities. \cite{sandratskii,jpcm}

The paper is organized as follows: we first present the details of the method, our
numerical implementation and numerical tests of one- and two-center integrals (section III).
We then demonstrate the capability of the proposed scheme with predictions for
group {IV} and {III-V} semiconductors and for 5$d$ metals (sections IV and V, respectively).

%
%

\section{The On-site approximation}

\label{meth}
\subsection{Relativistic effects in pseudopotential methods}

Kleinman and Bylander have shown that the procedures for generation of non-relativistic pseudopotentials can
be easily extended to account for relativistic effects. \cite{kleinman,bachelet} This relies
on solving self-consistently the all-electrons Dirac equation for a single atom and in the extraction of
a  pseudopotentials $V_j$, where now $j$ is the total angular momentum $j=l\pm\frac{1}{2}$.
The pseudopotential Hamiltonian can therefore be written as
\begin{equation}
\label{potrel}
\hat{V}=\sum\limits_{j,m_j} | j,m_j\ket\, V_j \,\bra j,m_j |,
\end{equation}

and includes all relativistic corrections to order $\alpha^2$, where $\alpha$ is the
fine structure constant and $| j,m_j\ket$ are total angular momentum states.
This expression can be recast in a form suitable for
non-relativistic pseudopotential theory by expressing
$| j,m_j\ket$ in terms of a tensor product of the regular angular momentum
states $| l,m\ket$ and the eigenstates of the $z$ component of the Pauli spin
matrices\cite{schwabl}

\begin{eqnarray}
| j=l+\frac{1}{2},m_j\ket&=&
\left(\begin{array}{c}
\sqrt{\frac{l+m_j+\frac{1}{2}}{2l+1}}\,\,| l,m_j-\frac{1}{2}\ket\\
\sqrt{\frac{l-m_j+\frac{1}{2}}{2l+1}}\,\,| l,m_j+\frac{1}{2}\ket
\end{array}\right),\nonumber\\
| j= l-\frac{1}{2},m_j\ket&=&
\left(\begin{array}{c}
\sqrt{\frac{l-m_j+\frac{1}{2}}{2l+1}}\,\,| l,m_j-\frac{1}{2}\ket\\
-\sqrt{\frac{l+m_j+\frac{1}{2}}{2l+1}}\,\,| l,m_j+\frac{1}{2}\ket
\end{array}\right).
\end{eqnarray}
Equation (\ref{potrel}) can then be rewritten\cite{units} as

\begin{eqnarray}
\label{pot_nrel}
\hat{\mathbf{V}} & = &
\hat{\mathbf{V}}^{sc}+\hat{\mathbf{V}}^{SO}=\nonumber\\
& = &
\sum\limits_{l,m}\,[\,\bar{V}_l\,\openone_{\sigma}+
\bar{V}_l^{SO}\vec{L}\cdot\vec{\mathbf{S}}\,]
\,\,| l, m\ket\bra l, m|,
\end{eqnarray}

where we use bold letters to indicate 2x2 matrices, with $\openone_{\sigma}$
representing the unit operator in spin space,
\begin{equation}
\vec{L}\cdot\vec{\mathbf{S}}=\frac{1}{2}
\left(\begin{array}{cc}\hat{L}_z&\hat{L}_-\\\hat{L}_+&-\hat{L}_z
\end{array}\right),
\end{equation}
and,
\begin{eqnarray}
\bar{V}_l^{\phantom{SO}} & = &
\frac{1}{2l+1}[(l+1)V_{l+\frac{1}{2}}+lV_{l-\frac{1}{2}}],\nonumber\\
\bar{V}_l^{SO} & = &
\frac{2}{2l+1}[V_{l+\frac{1}{2}}-V_{l-\frac{1}{2}}].
\end{eqnarray}
It should be stressed that the scalar part $\hat{\mathbf{V}}^{sc}$ of the pseudopotential
contains now not only the conventional non-relativistic pseudopotential, but also the scalar
relativistic corrections.

The vectors $| l, m\ket$, representing complex spherical harmonics,
form a complete  basis for the Hilbert space of the angular momentum operator
$\vec{L}$. It is a useful practice in solid state physics and quantum chemistry
to replace them with real spherical harmonics $| l, M\ket$, since the corresponding
Hamiltonian is a real matrix. The change of basis is achieved by the following
unitary transformation
\begin{eqnarray}
| l, M \ket & = &
\frac{1}{\sqrt{2}}\,\left(\,| l, m\ket + (-1)^m\,| l, \overline{m}\,\ket \right),\nonumber\\
| l,\overline{M}\,\ket & = &
\frac{1}{\sqrt{2}\,i}\,\left(\,| l, m\ket - (-1)^m\,| l, \overline{m}\,\ket \right),
\end{eqnarray}
which is valid for $M>0$ ($\,\overline{M}=-M$, $\overline{m}=-m$). The case $M=0$ is simply
given by $| l, M=0\ket = | l, m=0 \ket$.

The pseudopotential operator $\hat{\mathbf{V}}$ of equation (\ref{pot_nrel}), is now written as
\begin{eqnarray}
\hat{\mathbf{V}} & = &
\hat{\mathbf{V}}^{sc}+\hat{\mathbf{V}}^{SO} = \nonumber\\
& = &\sum\limits_{l,M}[\,\bar{V}_l\,\openone_{\sigma}+%
\bar{V}_l^{SO}\vec{L}\cdot\vec{\mathbf{S}}\,]%
| l, M\ket\bra l, M|\;.
\end{eqnarray}

Finally the Kohn-Sham Hamiltonian \cite{LDA} is expressed as a sum
of kinetic energy $\hat{\mathbf{T}}$, scalar relativistic $\hat{\mathbf{V}}^{sc}$,
spin-orbit $\hat{\mathbf{V}}^{SO}$, Hartree $\hat{\mathbf{V}}^{H}$ and exchange
and correlation $\hat{\mathbf{V}}^{xc}$ potentials
\begin{equation}
\hat{\mathbf{H}}=\hat{\mathbf{T}}+\hat{\mathbf{V}}^{sc}+\hat{\mathbf{V}}^{SO}+
\hat{\mathbf{V}}^{H}+\hat{\mathbf{V}}^{xc}\;.
\end{equation}

This Hamiltonian is therefore a $2\times2$ matrix in spin space
\begin{equation}
\hat{\mathbf{H}}=\left[\begin{array}{cc}
\hat{H}^{\uparrow\uparrow}&\hat{H}^{\uparrow\downarrow}\\
\hat{H}^{\downarrow\uparrow}&\hat{H}^{\downarrow\downarrow}
\end{array}\right],
\end{equation}
whose non-diagonal blocks arise from the exchange and correlation potential
whenever the system under consideration displays a non-collinear spin, and
also from the spin-orbit potential.\cite{sandratskii,jpcm}

\subsection{Spin-orbit in LCAO schemes: the on-site approximation}
LCAO methods expand the eigenstates $| \psi_n\ket$ of the non-collinear
Kohn-Sham Hamiltonian over a set of localised orbitals $| \phi_i\ket$,

 \begin{equation}
 | \mathbf{\psi}_n \ket=\sum_i
 \left(\begin{array}{c} c_{n,i}^{\uparrow}\\c_{n,i}^{\downarrow}\end{array}\right)
 \,\,|\phi_i\ket
 \end{equation}

where $i$ is a collective index for all the symbols required to describe
uniquely a given orbital
\begin{equation}
|\phi_i\ket=| \phi_{n_i,l_i,M_i}\ket=| R_{n_i,l_i}\ket\otimes \,| l_i, M_i\ket\:.
\end{equation}
Here both the radial and angular part of the wave function
$\phi_i(\vec{r}-\vec{d}_i)=\bra \vec{r}\ |\phi_i\ket$ is centered at the position $\vec{d}_i$.
Note that $n_i$ does not necessarily describe the principle quantum number only, but
generally labels a set of radial functions associated to the angular momentum $l_i$
according to the multiple zetas scheme.\cite{PAO}

The Kohn-Sham equation, $\hat{H}\,| \psi_n \ket = E_n\,| \psi_n\ket$, is then
projected onto such orbital basis set as
\begin{equation}
\left[\begin{array}{c c}
H^{\uparrow\uparrow}_{ij}-E_n S_{ij} &H^{\uparrow\downarrow}_{ij}\\
H^{\downarrow\uparrow}_{ij} & H^{\downarrow\downarrow}_{ij}-E_n S_{ij}
\end{array}\right]
\,
\left[\begin{array}{c}
c_{n,j}^\uparrow\\
c_{n,j}^\downarrow
\end{array}\right]=0,
\end{equation}
%
where $H^{\sigma\sigma^\prime}_{ij}=\bra \phi_i | \hat{H}^{\sigma\sigma^\prime} | \phi_j\ket$
and $S_{ij}=\bra \phi_i | \phi_j\ket$ are the matrix elements of the Hamiltonian and overlap matrix
respectively.

The spin-orbit term can then be calculated as
\begin{eqnarray}
\label{sss}
\mathbf{V}^{SO}_{ij} & = &
\bra \phi_i | \hat{\mathbf{V}}^{SO} | \phi_j\ket = \nonumber\\
& = & \!\!\!\!\sum\limits_{k,l_k,M_k}\!\!\!%
\bra \phi_i | \bar{V}_{l_k}^{SO}\vec{L}\cdot\vec{\mathbf{}S}\,| l_k,%
M_k\ket\,\bra l_k, M_k| \phi_j\ket,%
\end{eqnarray}
where index $k$ indicates the atom on which the potential is
centered, $\bar{V}_{l_k}^{SO}=\bar{V}_{l}^{SO}(\vec{r}-\vec{d}_k)$ and
$\Ket{l_k,M_k}$ is the spherical harmonic centered at the same atomic position
${\vec d}_k$.
Equation (\ref{sss}) involves a considerable
number of three-center integrals. Inclusion of the
spin-orbit interaction is therfore, in the LCAO approach, if straightforward,
computationally intensive. One possibility of reducing the computational effort
is to construct fully non-local pseudopotentials. \cite{kleinman:1425}
We note however that the radial part of the spin-orbit pseudopotentials,
$\bar{V}^{SO}_l$, is very short-ranged, resulting in
matrix elements that decay quickly with the distance
among the three centers. Thus we consider only matrix elements where
the three components, both orbitals and the pseudopotential, reside on
the same atom, and discard all the rest. This approximation simplifies the
matrix elements of equation (\ref{sss}) to one center integrals
and drastically reduces the computational effort needed to account
for spin-orbit effects. Then our approximated matrix elements are
\begin{eqnarray}
\fl {\mathbf{V}}^{SO}_{ij} &=&
\frac{1}{2}\sum\limits_{k,l_k>0,M_k}
\bra R_{n_i,l_i} | \bar{V}_{l_k}^{SO} | R_{n_j,l_j}\ket\,
\bra l_i, M_i | \left(\begin{array}{cc}\hat{L}_z&\hat{L}_-\\\hat{L}_+&-\hat{L}_z
\end{array}\right)\,| l_k, M_k\ket\,\bra l_k, M_k| l_j, M_j\ket\nonumber\\
&\approx&
\frac{1}{2}\,
\bra R_{n_i,l_i} | \bar{V}_{l_i}^{SO} | R_{n_j,l_i}\ket\,
\bra l_i, M_i | \left(\begin{array}{cc}\hat{L}_z&\hat{L}_-\\\hat{L}_+&-\hat{L}_z
\end{array}\right)\,| l_i, M_j\ket \,\delta_{l_i,l_j},
\end{eqnarray}
since the $\hat{L}_\alpha$ operators leave each $l_i$ subspace invariant.
The angular part of these on-site matrix elements can be calculated analytically\footnote{
This formula was not correctly written in the previous version of the manuscript. We wish to
thank Hyungjun Lee from Yonsei University for kindly drawing our attention to this point.
The correct formulae were in any case used in the code from the very beginning and, hence
all the results of the simulations presented in the article are correct and remain the same.}
%
\begin{eqnarray}
\fl\bra l, M_i     | \hat{L}_z     | l, M_j \ket &=& - i M_i \delta(M_i+M_j = 0) \\ \nonumber
\fl\bra l, 0       | \hat{L}_\mp   | l, M_j \ket   &=& \pm \sqrt{\frac{l(l+1)}{2}} \delta(M_j=1) - i \sqrt{\frac{l(l+1)}{2}} \delta(M_j=\bar{1})\\ \nonumber
\fl\bra l, 1       | \hat{L}_\mp   | l, M_j \ket   &=& -i \frac{\sqrt{l(l+1)-2}}{2} \delta(M_j=\bar{2}) \mp \sqrt{\frac{l(l+1)}{2}} \delta(M_j=0) \pm \frac{\sqrt{l(l+1)-2}}{2} \delta(M_j=2) \\ \nonumber
\fl\bra l,\bar{1}  | \hat{L}_\mp   | l, M_j \ket   &=&\pm \frac{\sqrt{l(l+1)-2}}{2} \delta(M_j=\bar{2}) + i \sqrt{\frac{l(l+1)}{2}} \delta(M_j=0) + i \frac{\sqrt{l(l+1)-2}}{2} \delta(M_j=2) \\ \nonumber
\fl\bra l, 2       | \hat{L}_\mp   | l, M_j \ket   &=& -i \frac{\sqrt{l(l+1)-6}}{2} \delta(M_j=\bar{3}) - i \frac{\sqrt{l(l+1)-2}}{2} \delta(M_j=\bar{1})
\\\nonumber & & \mp \frac{\sqrt{l(l+1)-2}}{2} \delta(M_j=1) \pm \frac{\sqrt{l(l+1)-6}}{2} \delta(M_j=3)\\ \nonumber
\fl\bra l,\bar{2}  | \hat{L}_\mp   | l, M_j \ket   &=&\pm \frac{\sqrt{l(l+1)-6}}{2} \delta(M_j=\bar{3}) \mp \frac{\sqrt{l(l+1)-2}}{2} \delta(M_j=\bar{1})
\\\nonumber & & +i \frac{\sqrt{l(l+1)-2}}{2} \delta(M_j=1) + i \frac{\sqrt{l(l+1)-6}}{2} \delta(M_j=3)\\ \nonumber
\fl\bra l, 3       | \hat{L}_\mp   | l, M_j \ket   &=& -i \frac{\sqrt{l(l+1)-12}}{2}\delta(M_j=\bar{4}) - i \frac{\sqrt{l(l+1)-6}}{2} \delta(M_j=\bar{2})
\\\nonumber & & \mp \frac{\sqrt{l(l+1)-6}}{2} \delta(M_j=2) \pm \frac{\sqrt{l(l+1)-12}}{2} \delta(M_j=4)\\ \nonumber
\fl\bra l,\bar{3}  | \hat{L}_\mp   | l, M_j \ket   &=&\pm \frac{\sqrt{l(l+1)-12}}{2}\delta(M_j=\bar{4}) \mp \frac{\sqrt{l(l+1)-6}}{2} \delta(M_j=\bar{2})
\\\nonumber & & +i \frac{\sqrt{l(l+1)-6}}{2} \delta(M_j=2) + i \frac{\sqrt{l(l+1)-12}}{2} \delta(M_j=4)\\ \nonumber
\end{eqnarray}

Some useful symmetries of the matrix elements of the Hamiltonian and of its spin-orbit
part should also be highlighted. Since the Hamiltonian is hermitian the
matrix elements satisfy the relation
\begin{equation}
H_{ij}^{\sigma\sigma^\prime}=(H_{ji}^{\sigma^\prime\sigma})^*.
\end{equation}
Moreover it is also easy to show that all the terms in the Hamiltonian satisfy a spin box
hermiticity, {\it i.e.}
\begin{equation}
H_{ij}^{\sigma\sigma^\prime}=(H_{ij}^{\sigma^\prime\sigma})^*,
\end{equation}
except for the spin-orbit contribution which is spin box anti-hermitian
\begin{equation}
H_{ij}^{\sigma\sigma^\prime}=-(H_{ij}^{\sigma^\prime\sigma})^*.
\end{equation}
This property has important consequences for the calculation of the total energy.

\subsection{Density matrix and total energy}
The charge density can also be written in terms of the LCAO basis as
\begin{eqnarray}
\mathbf{n}(\vec{r}) & = &
\sum_n\,\,f_n\,\mathbf{\psi}_n(\vec{r})\,\,\mathbf{\psi}_n(\vec{r})^\dagger =\nonumber\\
& = & \sum_{i,j}
\phi_i(\vec{r}-\vec{d}_i)\,\phi_j^*(\vec{r}-\vec{d}_j)\,\,\mathbf{\rho}_{ij}
\end{eqnarray}
where $f_n$ represents the occupation of the Kohn-Sham eigenstate $\psi_n(\vec{r}\,)$
and $\mathbf{\rho}_{ij}$ is a $2\times 2$ matrix containing the products of wave-function
coefficients, whose components are
\begin{equation}
\rho_{ij}^{\sigma\sigma^\prime}=\sum\limits_n \,f_n \,\,c_{n,i}^\sigma \,c_{n,j}^{\sigma^\prime*}.
\end{equation}

The electronic contribution to the total energy may be expressed as a sum of a band-structure (BS)
contribution plus double-counting corrections. The BS contribution can be written in the LCAO basis as
\begin{equation}
E_e^{BS}=\sum\limits_n f_n \,\,\bra \mathbf{\psi}_n | \hat{\mathbf{H}} | \mathbf{\psi}_n\ket
=\sum\limits_{i,j,\sigma,\sigma^\prime} H_{ij}^{\sigma\sigma^\prime}
\rho_{ji}^{\sigma^\prime\sigma},
\end{equation}
which may also be expressed as
\begin{eqnarray}
& & \sum\limits_{ij}%
\left\{ H_{ij}^{\uparrow\uparrow}\,\rho_{ji}^{\uparrow\uparrow}+
H_{ij}^{\downarrow\downarrow}\,\rho_{ji}^{\downarrow\downarrow}+\right.\nonumber\\
& & \left. + 2\,\Real\left[(V_{ij}^{xc \uparrow\downarrow}%
-V_{ij}^{SO \uparrow\downarrow})\,
(\rho_{ji}^{\uparrow\downarrow})^*\right]\right\}\:,
\end{eqnarray}
where we have isolated the non-diagonal contributions in spin space.
These arise only from the spin-orbit interaction and from the exchange and
correlation potential whenever spin non-collinearity is present.
The spin-orbit contribution to the total energy therefore is
\begin{equation}\label{eq:nrg-so}
E^{SO}= \Trace \sum_{i,j} \mathbf{V}^{SO}_{ij}\,
\mathbf{\rho}_{ji},
\end{equation}
since there are no double-counting terms.

\subsection{Forces and stresses}
One important consequence of the on-site approximation is that spin-orbit does not give rise to an explicit
contribution to forces and stresses, even though an implicit contribution due to the modification of the
self-consistent wave-function is always present.  According to
Hellmann-Feynman theorem, \cite{feynman:340} the spin-orbit contribution
to the forces excerpted upon an atom centered at position $\vec{d}_k$ is obtained by
simply differentiating the energy with respect to the atomic coordinates of the atom,
\begin{eqnarray}
-\vec{F}_k^{SO}&\!\!=\!\!&\vec{\nabla}_{k}E^{SO}=\nonumber\\
&\!\!=\!\!&\Trace\sum_{i,j}\!\!\left\{\!\left[\vec{\nabla}_{k}\hat{\mathrm{V}}^{SO}_{ij}\right]\!%
\mathbf{\rho}_{ji}\!+\!\mathbf{V}^{SO}_{ij}\!\!%
\,\left[\vec{\nabla}_{k}\hat{\mathrm{\rho}}_{ji}\right]\!\right\}\label{forso},
\end{eqnarray}

where both the $i$ and $j$ orbitals are centered at the same atomic position $\vec{d}_k$
and $\vec{\nabla}_{k}=\vec{\nabla}_{\vec{d}_k}$.
However both contributions to the spin-orbit forces in equation (\ref{forso}) vanish.
The first one is identically zero since the one-center integrals do not depend on the atomic position,
\begin{equation}\label{eq:grad_Hso}
\nabla_{k} \bra R_{n_i,l_i} | \bar{V}_{l_k}^{SO} | R_{n_j,l_j}\ket \equiv 0\:.
\end{equation}

The second term may be rewritten as
\begin{equation}
-\sum_{ij} \left(\mathcal{E}^{SO,\uparrow\uparrow}_{ij}+
\mathcal{E}^{SO,\downarrow\downarrow}_{ij} \right) \vec{\nabla}_k {S}_{ji},
\end{equation}
where $\mathcal{E}^{SO,\sigma\sigma'}_{ij}$ are the components of the spin-orbit contribution
to the energy-density matrix
\begin{equation}
\mathcal{E}_{ij}^{SO} =\frac{1}{2}\,\sum\limits_{l,m}\,(S_{il}^{-1}\,\mathbf{V}^{SO}_{lm}\,\mathbf{\rho}_{mj}+
\mathbf{\rho}_{il}\,\mathbf{V}^{SO}_{lm}\,S_{mj}^{-1}).
\end{equation}
However, since $\mathcal{E}^{SO}$ is antisymmetric with respect to the
orbital indices, in contrast to the overlap matrix that is symmetric,
the second term vanishes as well.

In a similar way one can demonstrate that the spin-orbit interaction in the on-site approximation
does not introduce any additional contribution to the stress.

%
%


\section{Numerical tests on one- and two-center integrals}

The validity of the on-site approximation relies on the fact that two and three center
integrals are much smaller than the one-center ones, which are kept in  the
simulation. Among those two and three center matrix elements the two-center
integrals
\begin{equation}
\mathbf{V}^{2c}_{ij}(\vec{R}) =
\bra \phi_i (\vec{R})\,| \bar{V}_{l_j}^{SO}(0)\,\vec{L}(0)\cdot\vec{\mathbf{}S}(0)\,|
\phi_j(0)\ket,
\end{equation}
are expected to have the largest absolute value. An excellent test consists of
calculating these matrix elements for a given material along the direction
that joins one atom with its nearest neighbours, as a function of the
distance $R$ between the two centers. Then, $R=0$ provides the on-site matrix
elements, while if $R$ equals the nearest neighbours distance, the calculation
describes the desired two-center integrals. We have performed such test
for a representative semiconductor, Si, and a representative 5$d$ metal, Pt.

\begin{figure}[t]
\begin{center}
\includegraphics[width=6cm,angle=-90]{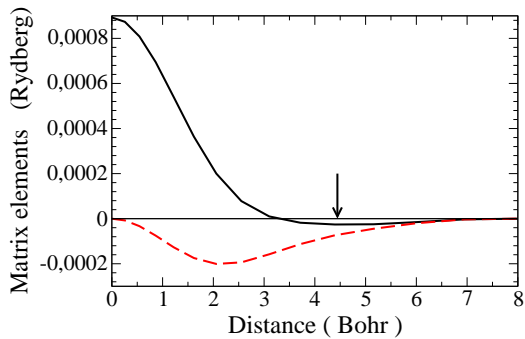}
\end{center}
\caption{Two-center integrals $\mathbf{V}_{y-x}^z(R)$ and $\mathbf{V}_{z-x}^z(R)$
(solid and dashed lines, respectively) for silicon along the (111) direction,
as a function of the distance R between the two centers. The arrow indicates
the distance of the nearest-neighbour atoms.}
\label{Si-2C}
\end{figure}

For silicon, the valence electrons include s- and p-orbitals only and therefore we consider the
following matrix elements
\begin{eqnarray}
\mathbf{V}_{y-x}^z(R) &=&
\bra \phi_{p_y} (R)\,| \bar{V}_{p}^{SO}(0)\,L_z(0)\,|\phi_{p_x}(0)\ket\nonumber\:,\\
\mathbf{V}_{z-x}^z(R) &=&
\bra \phi_{p_z} (R)\,| \bar{V}_{p}^{SO}(0)\,L_z(0)\,|\phi_{p_x}(0)\ket \:.
\end{eqnarray}
For $R$ =0, the first of these matrix elements reduces to the on-site integral
$\mathbf{V}_{y-x}^z(0)$, while the second is zero by symmetry. Fig. \ref{Si-2C} shows
the matrix elements as a function of R along the direction (111). The matrix elements,
evaluated at the nearest neighbour distance are considerably smaller than the on-site
integral, with the $\mathbf{V}(R)/\mathbf{V}(0)$ ratio being $\sim$0.03 and $\sim$0.08
respectively for $y-x$ and $z-x$.

In the case of Pt we have considered not only the 5$d$ but also the 6$p$ orbitals. Platinum
crystalizes with an fcc structure and a nearest neighbours distance of 5.4 a.u.
We compute the same $p$-matrix elements as for silicon, as well as the two
following $d$ integrals
\begin{eqnarray}
\mathbf{V}_{x^2y^2-xy}^z(R) &=&
\bra \phi_{x^2-y^2} (R)\,| \bar{V}_{d}^{SO}(0)\,L_z(0)\,|\phi_{xy}(0)\ket\:,\nonumber\\
\mathbf{V}_{z^2-xz}^z(R) &=&
\bra \phi_{3z^2-r^2} (R)\,| \bar{V}_{d}^{SO}(0)\,L_z(0)\,|\phi_{xz}(0)\ket\:.
\end{eqnarray}

\begin{figure}[t]
\begin{center}
\includegraphics[width=6cm,angle=-90]{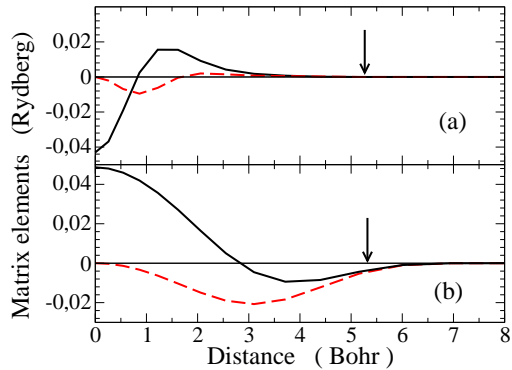}
\end{center}
\caption{(a) Two-center integral $\mathbf{V}_{x^2-y^2-xy}^z(R)$
$\mathbf{V}_{z^2-xz}^z(R)$ (solid and dashed lines, respectively) and
(b) $\mathbf{V}_{y-x}^z(R)$ and $\mathbf{V}_{z-x}^z(R)$ (also, solid and dashed lines, respectively) for platinum
along the (110) direction as a function of the distance R between the two centers.
The arrow indicates the distance of the nearest-neighbour atoms.}
\label{Pt-2C}
\end{figure}

Fig. \ref{Pt-2C} shows these matrix elements as a function of $R$. For the $d$-type matrix elements
(figure \ref{Pt-2C}a) the decay with the distance between the centers is rather dramatic
and the two-center matrix elements are about $10^{-4}$ times smaller than the on-site integrals.
The case of the $p$-integrals (figure \ref{Pt-2C}b) is similar to that of Si with a $\mathbf{V}(R)/\mathbf{V}(0)$ ratio
of $\sim 0.08$. For the specific case of Pt however these $p$-integrals are expected to contribute little to
any physical quantities since they correspond to unoccupied states.

Same tests for other materials give similar results. We therefore conclude that the on-site approximation provides
accurate results for heavy transition metals and good quantitative estimates for semiconductors. For this latter we
estimate an error due to the neglecting of high-center integrals not larger than 10\%, and on average of the order
of 3-6\%.

%
%


\section{Spin-orbit in group IV and III-V semiconductors}
\label{res}


We have thus demonstrated that in general our on-site approximation
simplifies the inclusion of spin-orbit effects in LCAO DFT codes. We have thus
implemented the method in SIESTA, a LCAO code able to simulate non-collinear
arrangements of spins,\cite{siesta,jpcm} that in addition reads
relativistically generated pseudopotentials in the form required by equation
(\ref{potrel}).

The numerical implementation is rather simple since the spin-orbit
contribution to the Hamiltonian does not depend on the electron charge
density and therefore does not need to be updated during the self-consistent
procedure. This drastically reduces the computational overheads, which
are almost identical to those of a standard non-collinear spin-polarised calculation.
Here we present a series of test cases for the band-structures of
group IV and III-V semiconductors, obtained with the local spin density
approximation and norm-conserving relativistic pseudopotentials.
Special care was taken in the generation of the pseudopotentials and
of the basis sets and in the choice of the parameters that control the numerical
accuracy of real and reciprocal space integrals.\cite{siesta}
In particular the basis set was highly optimised following the scheme proposed in
references [\cite{basis1,basis2}], from which we have borrow our
notation for multiple zeta basis sets SZ, SZP, DZ, DZP, TZ,
TZP, TZDP, TZTP, TZTPF.

The introduction of the spin-orbit interaction lifts specific degeneracies in the
band-structure of a material. In particular,
for diamond and zincblende semiconductors the six fold degenerate valence
band $\Gamma^v_{15}$, at $\Gamma$ splits in two. The first is four-fold degenerate
$\Gamma^v_8$ (the heavy and light hole bands), and the second is only two fold degenerate
$\Gamma^v_7$ (the spin-split-off band). This energy splitting
$\Delta_0\ (\Gamma^v_{15}\to\Gamma^v_8,\Gamma^v_7)$, is the hallmark of the effects of
spin-orbit interaction in the band-structure of these materials. Other commonly measured energy splittings
are called $\Delta^\prime_0\ (\Gamma^c_{15}\to\Gamma^c_8,\Gamma^c_7)$,
$\Delta_1\ (L^v_3\to L^v_{4,5},L^v_6)$ and $\Delta^\prime_1\ (L^c_3\to L^c_{4,5},L^c_6)$.
\begin{figure}[t]
\begin{center}
\includegraphics[height=5.6cm,width=8cm,angle=-90]{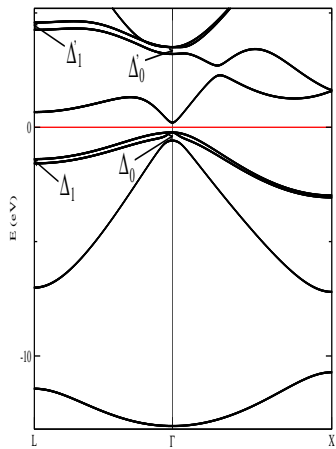}
\end{center}
\caption{\label{gaas}Band structure of GaAs calculated with the on-site approximation to
the spin-orbit interaction.}
\end{figure}

We show in Fig. \ref{gaas} the band-structure of the canonical III-V semiconductor, GaAs. The figure
also defines graphically the energy splittings described in the previous paragraph. We note first that
the band-structure closely resembles that obtained with other methodologies, and also
agrees rather well with the experimental data,\cite{wepfer} except for the characteristic LDA
underestimation of the semiconductor gap, which is further reduced because of the spin-orbit energy splitting.
Therefore, in all the tests that follow, we have avoided narrow gap semiconductors, which are usually
predicted to be metals by LDA.\cite{surh}
\subsection{Band structure of Si}
\begin{figure}[t]
\begin{center}
\includegraphics[height=9cm,width=10cm,angle=-90]{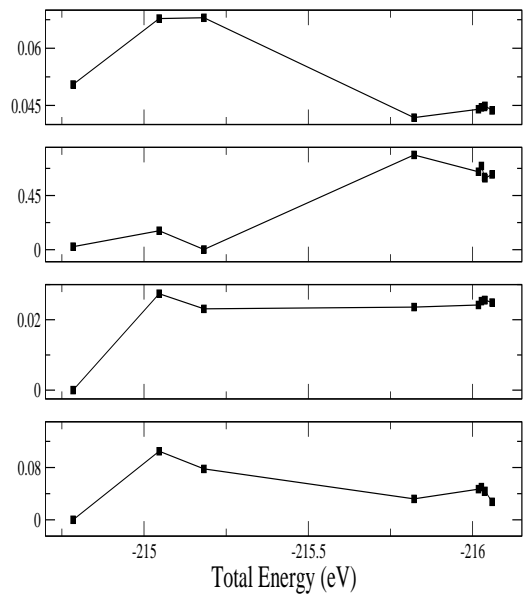}
\end{center}
\caption{\label{siconv}Convergence of the spin-orbit energy splittings of bulk Si with the
size of the basis set. From top to bottom panels, we present respectively $\Delta_0$, $\Delta^\prime_0$,
$\Delta_1$ and $\Delta^\prime_1$ as functions of the total energy associated with each set. The points
correspond to SZ, DZ, TZ, SZP, DZP, TZP, TZDP, TZTP and TZTPF, respectively.}
\end{figure}

The spin-orbit energy splittings of silicon at the high symmetry points of the diamond lattice are
presented in Table \ref{sisp} and Fig. \ref{siconv} for increasingly complete basis sets.
\begin{table}[t]
\begin{center}
\begin{tabular}{|c||c|c|c|c|c|}
\hline
Basis & $E_T$ (eV) & $\Delta_0$ (eV) & $\Delta_0^\prime$ (eV)& $\Delta_1$ (eV) & $\Delta^\prime_1$ (eV) \\ \hline
SZ   & -214.8 & 0.051 & 0.025 &   -   &   -   \\ \hline
DZ   & -215.0 & 0.068 & 0.157 & 0.027 & 0.105 \\ \hline
TZ   & -215.2 & 0.068 & 0.002 & 0.023 & 0.078 \\ \hline
SZP  & -215.8 & 0.042 & 0.787 & 0.024 & 0.032 \\ \hline
DZP  & -216.0 & 0.044 & 0.647 & 0.024 & 0.047 \\ \hline
TZP  & -216.0 & 0.045 & 0.696 & 0.025 & 0.050 \\ \hline
TZDP & -216.0 & 0.045 & 0.604 & 0.026 & 0.043 \\ \hline
TZTP & -216.0 & 0.045 & 0.593 & 0.026 & 0.044 \\ \hline
TZTPF& -216.1 & 0.044 & 0.615 & 0.025 & 0.030 \\ \hline
REF  &   -    & 0.044 & 0.04  & 0.02  & 0.03  \\ \hline
\end{tabular}
\end{center}
\caption{\label{sisp}Spin-orbit energy splittings of bulk Si calculated for increasingly
complete basis sets.
$E_T$ is the total energy, $\Delta_0$, $\Delta_0^\prime$, $\Delta_1$ and
$\Delta^\prime_1$ are the splittings as defined in the text. REF
corresponds to the reference values, experimental whenever possible, as
described in the literature.\cite{wepfer}}
\end{table}

Our results are in extremely good agreement with the theoretical and experimental data
available in the literature.\cite{wepfer} We note that a DZP basis set, which is
usually assumed to be the minimal basis needed to obtain reasonably converged results,
already provides accurate predictions for the energy splittings.
The $\Delta^\prime_0$ split is somehow an anomaly, and in general we find that
the splittings of the conduction bands are not as well described as those of the valence bands.

We note that Kohn-Sham eigenvalues should not be associated to single particle
excitation energies, since the former are simply the Lagrangian multipliers leading to
the Kohn-Sham equations. This is valid for both the valence and the conduction
bands. However it is commonly accepted that DFT band structures are a good first
approximation to single particle energies of occupied states. The conduction
bands are somehow different since such states do not contribute to either the total
energy and the density matrix, and therefore are irrelevant in DFT. For this reason
a stark disagreement in the conduction band splitting should not be surprising.
Moreover, the systematic overestimation of the $\Delta^\prime_0$ splitting is related to
the underestimation of the bandgap. This produces an erroneous enhancement of the
hybridisation between the orbitals forming the conduction bands with those
forming the valence one, with a net overestimation of the spin-orbit splitting.
It is therefore expected that corrective schemes to the bandgap may also correct
the spin-orbit splitting of the conduction bands.

It is also important to note that basis where the radial component varies sharply around the
origin should be avoided. These in fact are difficult to integrate in the range where the
spin-orbit pseudopotential is appreciable and brings considerable numerical instability to
the evaluation of the matrix elements.

Finally we have calculated the Si bulk parameters for different basis sets
in order to check that the inclusion of spin-orbit interaction does not change
significantly the LDA results. A
summary of all the computed structural parameters is presented in Table \ref{sibm}.
\begin{table}[t]
\begin{center}
\begin{tabular}{|c||c|c|c|}
\hline
Basis & $a$ (\AA) & $BM$ (GPa) & $E_c$ (eV)\\ \hline
DZP & 5.392 & 98.2 & 5.480\\ \hline
TZP & 5.389 & 98.8 & 5.485\\ \hline
TZTP& 5.388 & 98.2 & 5.491\\ \hline
PW  & 5.384 & 95.9 & 5.369\\ \hline
LAPW& 5.41  & 96   & 5.28\\ \hline
EXP & 5.43  & 98.8 & 4.64\\ \hline
\end{tabular}
\end{center}
\caption{\label{sibm}Structural parameters of bulk Si for several
basis sets. $a$ is the lattice parameter, $BM$ the bulk
modules and $E_c$ the cohesive energy. PW refers to a 50 Ryd cutoff plane-wave
calculation.\cite{siesta}
LAPW corresponds to an all-electrons linear-augmented-plane-wave calculation\cite{lapw}
and EXP to the experimental values.\cite{kittel}}
\end{table}


\subsection{III-V semiconductors}
We have further tested our method by calculating the various energy splittings of
several group IV and III-V semiconductors such as Ge, GaAs, AlAs and AlSb, i.e. of those
with a reasonably large bandgap. For all of them we have found again that a DZP basis set
provides essentially converged results. This is demonstrated in table \ref{sp} where we show
that the splittings calculated with a DZP basis agree rather well with other theoretical estimates
and with experimental values. Also in this case $\Delta_0^\prime$ is the exception for the same
reasons explained before.
\begin{table}[b]
\begin{center}
\begin{tabular}{|c||c|c|c|c|}
\hline
Material & $\Delta_0$ (eV) & $\Delta_0^\prime$ (eV) & $\Delta_1$ (eV) & $\Delta^\prime_1$ (eV)\\
\hline
Ge               & 0.2959 & 0.3783 & 0.1545 & 0.356 \\ \hline
REF\cite{aspnes} & 0.287  & 0.200  & 0.184  & 0.266 \\ \hline
GaAs             & 0.3573 & 0.3006 & 0.1857 & 0.319 \\ \hline
REF\cite{wepfer} & 0.34   & 0.26   & 0.23   & 0.11  \\ \hline
AlAs             & 0.3073 & 0.0762 & 0.1636 & 0.118 \\ \hline
REF\cite{zunger} & 0.28   & 0.04   & 0.20   &   -   \\ \hline
AlSb             & 0.6847 & 0.1752 & 0.3440 & 0.307 \\ \hline
REF\cite{wepfer} & 0.75   & 0.1    & 0.4    & 0.09  \\ \hline
\end{tabular}
\end{center}
\caption{\label{sp}Spin splittings for several III-V semiconductors as calculated
with a DZP basis set. REF correspond to reference values, experimental whenever
possible, as described in the literature.\cite{aspnes,wepfer,zunger}}
\end{table}

%
%

\section{5{\lowercase{\it d}} metals: A\lowercase{u} and P\lowercase{t}}

Since the spin-orbit interaction is a relativistic effect, it is expected to increase
with the atomic number. Therefore, metals from the fifth row of the periodic table are
an ideal test ground for our method. Among those, gold and platinum are
specially good candidates, since the first is a closed-$d$ shell noble metal while the
$d$-bands of the second have considerable weight at the Fermi energy. Moreover,
a number of {\it ab initio} spin-orbit calculations are
available \cite{corso:115106} which demonstrate that the spin-orbit interaction
is essential for the correct description of their band structures. To simulate
these two materials, we have again used the LDA approximation for exchange
and correlation potential, and constructed an optimised set consisting of two atomic
wave functions in each of the s-, p- and d-channels.
\begin{figure}
\begin{center}
\includegraphics[width=\columnwidth]{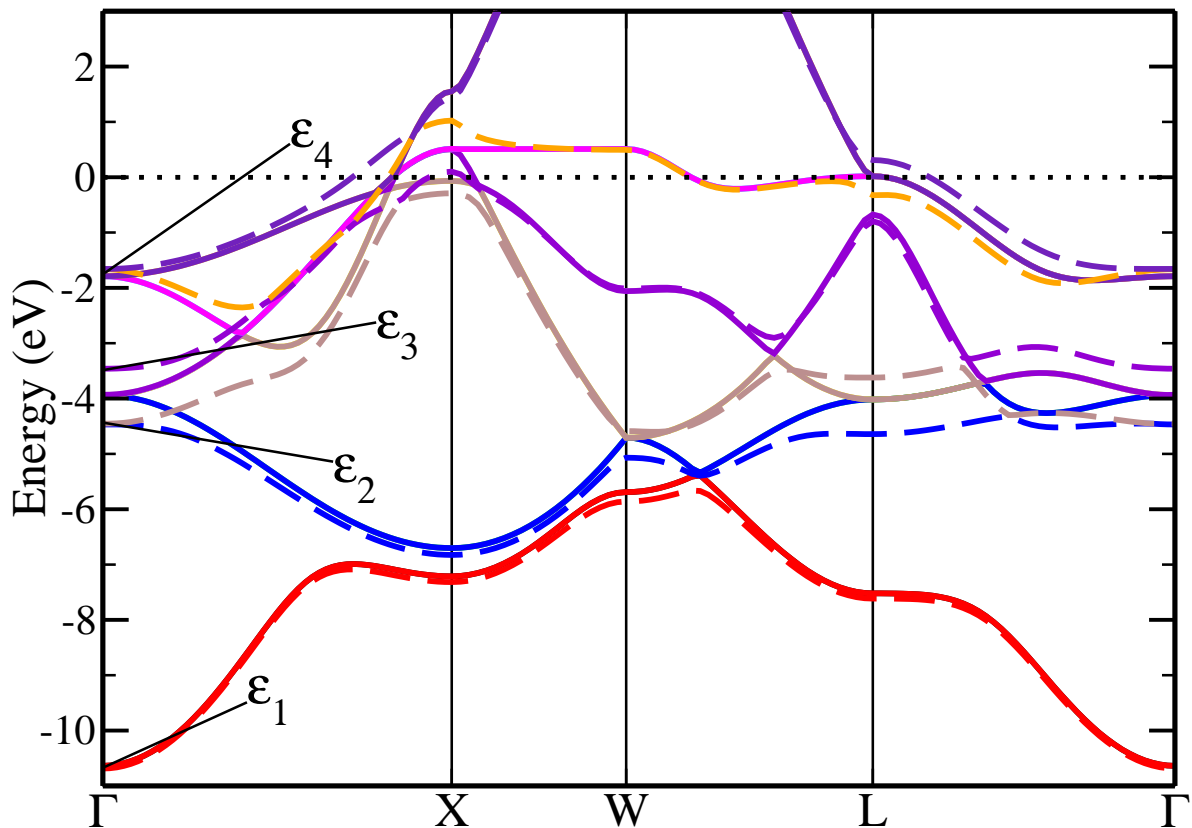}
\caption{Band-structure of platinum obtained at the theoretical equilibrium
lattice constant. The dashed line is for a spin-orbit calculation, while
the solid line is obtained when the spin-orbit coupling is not included. The figure
also provides a graphical definition of the energies
at the $\Gamma$ point, $\varepsilon_i$, of the bands that are closest to the Fermi energy.} \label{Pt-bands}
\end{center}
\end{figure}
\begin{figure}
\begin{center}
\includegraphics[width=\columnwidth]{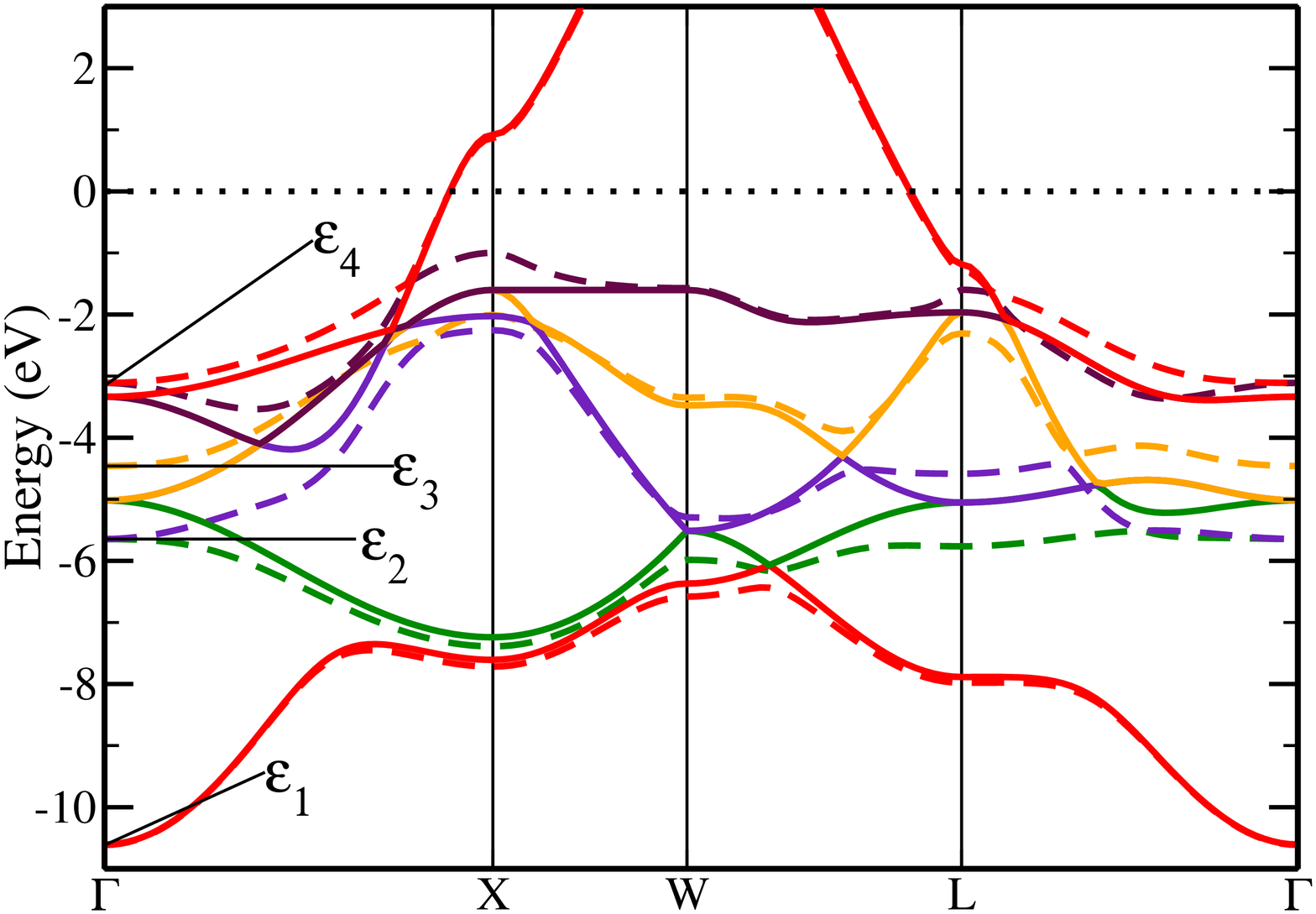}
\caption{Band structure of gold obtained at the equilibrium theoretical
lattice constant. The dashed line is for a spin-orbit calculation, while
the solid line is obtained when the spin-orbit coupling is not included.
The figure
also provides a graphical definition of the energies
at the $\Gamma$ point, $\varepsilon_i$, of the bands that are closest to the Fermi energy.
} \label{Au-bands}
\end{center}
\end{figure}

In Figs. \ref{Pt-bands} and \ref{Au-bands} we present the band-structures of Pt and Au,
calculated at the theoretical lattice constant, with and without taking into account the
spin-orbit coupling. the figures show that s-bands are not modified by
spin-orbit, while p- and d-bands suffer strong modifications, specially whenever
two bands cross each other. Moreover, we find that the spin-orbit interaction
lifts degeneracies at the band crossings as expected.

We summarise in Table \ref{table.Pt} the bulk lattice constant and the energy of some selected bands
at the $\Gamma$ point, that we define graphically in Fig. \ref{Pt-bands}.
These energies are in good agreement with other much more expensive methods, like
Plane-wave pseudopotential (PWSCF)\cite{corso:115106} and
relativistic full-potential Korringa-Kohn-Rostoker (KKR)\cite{theileis:13338} methods or
augmented plane-wave (APW) approaches for solving Dirac equation.\cite{kellen:11187}
We note that the differences range in the order of only a few per cent.

\begin{table}[t]
\begin{center}
\begin{tabular}{|l||r|r|r|r|}
\hline
Pt  & SIESTA+On-site & PWSCF &KKR &  APW\\ \hline
$a_0$ (Bohr)       &  7.45  &  7.40    &  7.36  &   7.414 \\ \hline
$\varepsilon_1$(eV)   & -10.53 &  -10.45  & -10.56 & -10.35  \\ \hline
$\varepsilon_2$(eV)   &  -4.38 &   -4.38  &  -4.48 & -4.24   \\ \hline
$\varepsilon_3$(eV)   &  -3.38 &   -3.39  &  -3.47 & -3.28   \\ \hline
$\varepsilon_4$(eV)   & -1.62  &   -1.53  &  -1.65 & -1.48  \\ \hline \hline
Au  &SIESTA+On-site & PWSCF &
KKR &  APW\\ \hline
$a_0$ (Bohr)       & 7.61 & 7.64 & 7.637 & 7.638 \\ \hline
$\varepsilon_1$(eV)   & -10.613 & -10.18 & -10.23 & -9.95 \\ \hline
$\varepsilon_2$(eV)   & -5.645  & -5.43 & -5.41 & -5.32\\ \hline
$\varepsilon_3$(eV)   &  -4.460 & -4.23 &  -4.23 & -4.14  \\ \hline
$\varepsilon_4$(eV)   & -3.111 &-2.97 & 3.04 & 2.96 \\ \hline
\end{tabular}
\end{center}
\caption{Lattice constant and energy $\varepsilon_i$ of selected bands at $\Gamma$ point obtained with the
use of the approximation presented in this article (SIESTA+On-site), and with PWSCF\cite{corso:115106},
KKR \cite{theileis:13338} and APW\cite{kellen:11187} methods. The definition of the energies $\varepsilon_i$
is provided in Figs. (3) and (4).}\label{table.Pt}
\end{table}


%


%



\section{Conclusion}

\label{conc}
We have presented a simple and effective method for including the spin-orbit
interaction in standard LCAO DFT calculations, which is based on relativistic
norm-conserving pseudopotentials and on an on-site approximation to the spin-orbit
matrix elements. The method is computational undemanding and extremely simple
to implement in standard LCAO DFT codes, such as SIESTA. Importantly the on-site
approximation does not introduce additional contributions to both forces and stress.
We have then presented a series of tests for group IV and III-V semiconductors
and for 5{\it d} metals. The overall structural parameters do not change with respect to
standard non-relativistic LDA calculations, and are in general good agreement with
reference data. The spin-orbit splittings of the band structures are also in
good agreement with those obtained with more computationally intensive DFT
methods.

The good results obtained for the electronic structures and structural parameters make
our method very attractive for large scale simulations where spin-orbit coupling is relevant.
This is for instance the case of semiconductors heterostructures and quantum-transport
calculations \cite{natmat} where spin-mixing effects are important.

\ack
We wish to thank D. S\'anchez-Portal, V. Garc\'{\i}a-Su\'arez, P. Ordej\'on and J. Soler for
useful discussions.
LFS acknowledges the support of FICYT through grant BP04-087. The research presented
in this article has been performed under the financial support of Science Foundation of
Ireland (SFI02/IN1/I175) and MEC through projects BFM2003-03156. Traveling has been
supported by Enterprise Ireland (IC 2004 84).



\section*{References}
\bibliographystyle{plain}

\begin{thebibliography}{10}

\bibitem{units}
Atomic units are used throughout the article.

\bibitem{natmat}
Steve W.~Bailey Alexandre R.~Rocha, Víctor M. García-Suárez, Colin~J. Lambert,
  Jaime Ferrer, and Stefano Sanvito.
\newblock Towards molecular spintronics.
\newblock {\em Nature Materials}, 4(4):335--339, 2005.

\bibitem{basis2}
Eduardo Anglada, Jos{\'e}~M. Soler, Javier Junquera, and Emilio Artacho.
\newblock Systematic generation of finite-range atomic basis sets for
  linear-scaling calculations.
\newblock {\em Physical Review B}, 66(205101):1--4, 2002.

\bibitem{aspnes}
D.~E. Aspnes.
\newblock Schottky-barrier electroreflectance of ge: Nondegenerate and
  orbitally degenerate critical points.
\newblock {\em Physical Review B (Solid State)}, 12(6):2297--2310, 1975.

\bibitem{bachelet}
Giovanni~B. Bachelet and M.~Sch{\"u}ter.
\newblock Relativistic norm-conserving pseudopotentials.
\newblock {\em Physical Review B}, 25(4):2103--2108, 1982.

\bibitem{bruno}
P.~Bruno.
\newblock {\em Electronic Structure: Basic Theory and Practical Methods}.
\newblock Cambridge University Press, Cambridge, 2004.

\bibitem{suecos}
Till Burkert, Olle Eriksson, Peter James, Sergei~I. Simak, Borje Johansson, and
  Lars Nordstrom.
\newblock Calculation of uniaxial magnetic anisotropy energy of tetragonal and
  trigonal fe, co, and ni.
\newblock {\em Physical Review B (Condensed Matter and Materials Physics)},
  69(10):104426, 2004.

\bibitem{corso:115106}
Andrea~Dal Corso and Adriano~Mosca Conte.
\newblock Spin-orbit coupling with ultrasoft pseudopotentials: Application to
  au and pt.
\newblock {\em Physical Review B (Condensed Matter and Materials Physics)},
  71(11):115106, 2005.

\bibitem{conquest}
M.~J.~Gillan D.~R.~Bowler, R.~Choudhury and T.~Miyazaki.
\newblock Recent progress with large-scale ab initio calculations: the conquest
  code.
\newblock {\em Physica status solidi (b)}, 243(5):989--1000, 2006.

\bibitem{dasdatta}
Supriyo Datta and Biswajit Das.
\newblock Electronic analog of the electro-optic modulator.
\newblock {\em Applied Physics Letters}, 56(7):665--667, 1990.

\bibitem{kellen:11187}
S.~Bei der Kellen and A.~J. Freeman.
\newblock Self-consistent relativistic full-potential korringa-kohn-rostoker
  total-energy method and applications.
\newblock {\em Physical Review B (Condensed Matter)}, 54(16):11187--11198,
  1996.

\bibitem{gaussian}
MJ~Frisch et~al.
\newblock {\em Gaussian 03, Revision C.02}.
\newblock Gaussian, Inc., Wallingford, CT, 2004.

\bibitem{feynman:340}
R.~P. Feynman.
\newblock Forces in molecules.
\newblock {\em Physical Review}, 56(4):340--343, 1939.

\bibitem{lapw}
C.~Filippi, D.~J. Singh, and C.~J. Umrigar.
\newblock All-electron local-density and generalized-gradient calculations of
  the structural properties of semiconductors.
\newblock {\em Physical Review B}, 50(20):14947--14951, 1994.

\bibitem{jpcm}
V~M Garc\'{i}a-Su\'{a}rez, C~M Newman, C~J Lambert, J~M Pruneda, and J~Ferrer.
\newblock Optimized basis sets for the collinear and non-collinear phases of
  iron.
\newblock {\em Journal of Physics: Condensed Matter}, 16(30):5453--5459, 2004.

\bibitem{roberta}
Dante Gatteschi and Roberta Sessoli.
\newblock Quantum tunneling of magnetization and related phenomena in molecular
  materials.
\newblock {\em Angew. Chem. Int. Ed.}, 42(3):268--297, 2003.

\bibitem{orderN}
Stefan Goedecker.
\newblock Linear scaling electronic structure methods.
\newblock {\em Reviews of Modern Physics}, 71(4):1085--1123, 1999.

\bibitem{DFT}
P.~Hohenberg and W.~Kohn.
\newblock Inhomogeneous electron gas.
\newblock {\em Phys. Rev.}, 136(3B):B864--B871, Nov 1964.

\bibitem{basis1}
Javier Junquera, Daniel S{\'a}nchez-Portal {\'O}scar~Paz, and Emilio Artacho.
\newblock Numerical atomic orbitals for linear-scaling calculations.
\newblock {\em Physical Review B}, 64(235111):1--9, 2001.

\bibitem{spinhall}
Y.~K. Kato, R.~C. Myers, A.~C. Gossard, and D.~D. Awschalom.
\newblock {Observation of the Spin Hall Effect in Semiconductors}.
\newblock {\em Science}, 306(5703):1910--1913, 2004.

\bibitem{stime}
J.~M. Kikkawa and D.~D. Awschalom.
\newblock Lateral drag of spin coherence in gallium arsenide.
\newblock {\em Nature}, 397(6715):139--141, 1999.

\bibitem{kittel}
C.~Kittel.
\newblock {\em Introduction to Solid State Physics}.
\newblock Wiley, New York, 1986.

\bibitem{kleinman}
Leonard Kleinman.
\newblock Relativistic norm-conserving pseudopotential.
\newblock {\em Physical Review B}, 21(6):2630--2631, 1980.

\bibitem{kleinman:1425}
Leonard Kleinman and D.~M. Bylander.
\newblock Efficacious form for model pseudopotentials.
\newblock {\em Physical Review Letters}, 48(20):1425--1428, 1982.

\bibitem{LDA}
W.~Kohn and L.~J. Sham.
\newblock Self-consistent equations including exchange and correlation effects.
\newblock {\em Phys. Rev.}, 140(4A):A1133--A1138, Nov 1965.

\bibitem{zunger}
Kurt~A. M\"ader and Alex Zunger.
\newblock Empirical atomic pseudopotentials for alas/gaas superlattices,
  alloys, and nanostructures.
\newblock {\em Phys. Rev. B}, 50(23):17393--17405, Dec 1994.

\bibitem{martin}
Richard~M. Martin.
\newblock {\em Electronic Structure: Basic Theory and Practical Methods}.
\newblock Cambridge University Press, Cambridge, 2004.

\bibitem{turbomole}
Marco Häser an Hans~Horn Reinhart~Ahlrichs, Michael~Bär and Christoph Kölmel.
\newblock Electronic structure calculations on workstation computers: The
  program system turbomole.
\newblock {\em Chemical Physics Letters}, 162(3):165--169, 1989.

\bibitem{sandratskii}
L.~M. Sandratskii.
\newblock Noncollinear magnetism in itinerant-electron systems: theory and
  applications.
\newblock {\em Adv. Phys.}, 47(1):91--160, 1998.

\bibitem{PAO}
Otto~F. Sankey and David~J. Niklewski.
\newblock Ab initio multicenter tight-binding model for molecular-dynamics
  simulations and other applications in covalent systems.
\newblock {\em Physical Review B (Condensed Matter)}, 40(6):3979--3995, 1989.

\bibitem{schwabl}
F.~Schwabl.
\newblock {\em Quantum Mechanics}.
\newblock Springer, Berlin, 2002.

\bibitem{fireball}
Spencer~D. Shellman, James~P. Lewis, Kurt~R. Glaesemann, Krzysztof Sikorski,
  and Gregory~A. Voth.
\newblock Massively parallel linear-scaling algorithm in an ab initio
  local-orbital total-energy method.
\newblock {\em J. Comput. Phys.}, 188(1):1--15, 2003.

\bibitem{onetep}
Chris-Kriton Skylaris, Peter~D Haynes, Arash~A Mostofi, and Mike~C Payne.
\newblock Using onetep for accurate and efficient $\mathcal{O}(n) $ density
  functional calculations.
\newblock {\em Journal of Physics: Condensed Matter}, 17(37):5757--5769, 2005.

\bibitem{siesta}
Jos{\'e}~M. Soler, Emilio Artacho, Julian~D. Gale, Alberto Garcia~Javier
  Junquera, Pablo Ordej{\'o}n, and Daniel S{\'a}nchez-Portal.
\newblock The {SIESTA} method for {\it ab initio} order-{N} materials
  simulations.
\newblock {\em Journal of Physics: Condensed Matter}, 14:2745--2779, 2002.

\bibitem{DMS}
Nicola A.~Hill Stefano~Sanvito, Gerhard~Theurich.
\newblock {Density Functional Calculations for III-V Diluted Ferromagnetic
  Semiconductors: A Review}.
\newblock {\em Journal of Superconductivity}, 15(1):85--104, 2002.

\bibitem{surh}
Michael~P. Surh, Ming-Fu Li, and Steven~G. Louie.
\newblock Spin-orbit splitting of {GaAs} and {InSb} bands near $\gamma$.
\newblock {\em Physical Review B}, 43(5):4286--4294, 1991.

\bibitem{theileis:13338}
V.~Theileis and H.~Bross.
\newblock Relativistic modified augmented plane wave method and its application
  to the electronic structure of gold and platinum.
\newblock {\em Physical Review B (Condensed Matter and Materials Physics)},
  62(20):13338--13346, 2000.

\bibitem{wepfer}
G.~G. Wepfer, T.~C. Collins, and R.~N. Euwema.
\newblock Calculated spin-orbit splittings of some group iv, iii-v, and ii-vi
  semiconductors.
\newblock {\em Physical Review B (Solid State)}, 4(4):1296--1306, 1971.

\bibitem{spintronics}
S.~A. Wolf, D.~D. Awschalom, R.~A. Buhrman, J.~M. Daughton, S.~von Molnar,
  M.~L. Roukes, A.~Y. Chtchelkanova, and D.~M. Treger.
\newblock {Spintronics: A Spin-Based Electronics Vision for the Future}.
\newblock {\em Science}, 294(5546):1488--1495, 2001.

\bibitem{cardona}
P.~Y. {Yu} and M.~{Cardona}.
\newblock {\em {Fundamentals of Semiconductors: Physics and Materials
  Properties}}.
\newblock Berlin: Springer, 2005.

\end{thebibliography}

\end{document}